\documentclass{jfm}

\usepackage{graphicx}
\usepackage{t1enc,alltt,amssymb,amsmath,natbib}

\ifCUPmtlplainloaded \else
  \checkfont{eurm10}
  \iffontfound
    \IfFileExists{upmath.sty}
      {\typeout{^^JFound AMS Euler Roman fonts on the system,
                   using the 'upmath' package.^^J}%
       \usepackage{upmath}}
      {\typeout{^^JFound AMS Euler Roman fonts on the system, but you
                   dont seem to have the}%
       \typeout{'upmath' package installed. JFM.cls can take advantage
                 of these fonts,^^Jif you use 'upmath' package.^^J}%
      }
  \else
  \fi
\fi

\ifCUPmtlplainloaded \else
  \checkfont{msam10}
  \iffontfound
    \IfFileExists{amssymb.sty}
      {\typeout{^^JFound AMS Symbol fonts on the system, using the
                'amssymb' package.^^J}%
       \usepackage{amssymb}%

      }{}
  \fi
\fi

\ifCUPmtlplainloaded \else
  \IfFileExists{amsbsy.sty}
    {\typeout{^^JFound the 'amsbsy' package on the system, using it.^^J}%
     \usepackage{amsbsy}}
    {\providecommand\boldsymbol[1]{\mbox{\boldmath $##1$}}}
\fi

\newcommand\Rey{\mbox{\textit{Re}}}  % Reynolds number
\newsavebox{\astrutbox}
\sbox{\astrutbox}{\rule[-5pt]{0pt}{20pt}}

\begin{document}

\title{Stabilizing effect of flexibility in the wake of a flapping foil}
%\author[C. Marais, B. Thiria, J. E. Wesfreid and R. Godoy-Diana]{Catherine Marais, Benjamin Thiria, Jos\'e Eduardo Wesfreid and Ramiro Godoy-Diana}
\author{C. Marais, B. Thiria, J. E. Wesfreid and R. Godoy-Diana}
%\email{ramiro@pmmh.espci.fr}
\affiliation{{Physique et M\'ecanique des Milieux H\'et\'erog\`enes (PMMH)}\\
{CNRS UMR7636; ESPCI ParisTech; UPMC (Paris 6); Univ. Paris Diderot (Paris 7)}\\
{10, rue Vauquelin, F-75005 Paris, France}}

%\date{\today}

%\pacs{
%    47.32.C-   % Vortex dynamics (fluid flow)
%    87.85.gj    % Biomechanics. Movement and locomotion
%    06.30.Gv   % Velocity, measurement of    % Need more PACS numbers
%    47.15.Tr    % Laminar wakes
%}

%\keywords{flapping foils, flexibility, propulsive performance, fluid-structure interaction}

\maketitle

\begin{abstract}
The wake of a flexible foil undergoing pitching oscillations in a low-speed hydrodynamic tunnel is used to examine the effect of chord-wise foil flexibility in the dynamical features of flapping-based propulsion. We compare the regime transitions in the wake with respect to the case of a rigid foil and show that foil flexibility inhibits the symmetry breaking of the reverse B\'enard-von K\'arm\'an wake reported in the literature. A momentum balance calculation shows the average thrust to be up to three times greater for the flexible foil than for the rigid foil. We explain both of these observations by analyzing the vortex dynamics in the very near wake.

\end{abstract}

\section{Introduction}

In nature, wings and fins are compliant structures
\cite[see e.g.][and references therein]{daniel2002}. During locomotion, these flapping appendages endure
deformation, actively or passively, which has an important role in
the propulsive and maneuvering capabilities of swimming \cite[][]{fish1999} and flying
animals \cite[][]{wootton1992}. A vast amount of work studying the effect of
wing compliance in flapping wing systems has been reported in the
literature \cite[see][for a review]{shyy2010}, and it is clear that
the redistribution of aerodynamic forces resulting from wing
deformation can have a dramatic effect in performance. Chordwise flexibility in particular has been 
shown  to increase the propulsive efficiency in various configurations of flapping-based propulsion  \cite[see e.g.][]{katz1978,heathcote2007,thiria2010}, and a few fluid physics mechanisms behind the beneficial effects of passive compliance for propulsive performance have been identified \cite[][]{eldredge2008,spagnolie2010,ramananarivo2011}. In this paper we show that, aside from its well-documented effect on the thrust force, chordwise flexibility plays an important role in the stability of the propulsive wake. It has been shown in the literature that the mean propulsive jet produced by flapping motion can sometimes lose its symmetry giving rise to a deflected jet \cite[][]{jones1998,lewin2003,heathcote2007b,godoy-diana2008,godoy-diana2009,vonellenrieder2008,buchholz2008,cleaver2010,yu2012}. The mechanism of deflection is triggered by the wake vortices being shed as dipolar structures, instead of as the regular array of counter-rotating vortices of the reverse B\'enard-von K\'arm\'an pattern characteristic of propulsive wakes produced by flapping motion. For the case of a rigid
teardrop-shaped foil undergoing pitching oscillations, the phase space for the transition between different regimes has been well established in previous work \cite[][]{godoy-diana2008,godoy-diana2009,marais}. We examine here the effect of introducing flexibility to the problem, performing a detailed comparison of the vortex dynamics in the near wake of the flapping foil for two cases: the \emph{rigid} foil (our benchmark case), and a \emph{flexible} foil with the same shape but made of a compliant material. In addition to quantifying the increase in the thrust force observed when using a flexible foil, we demonstrate that flexibility inhibits the symmetry breaking of the propulsive wake.

% FIGURE 1
\begin{figure}
\centering
\includegraphics[height=0.43\linewidth]{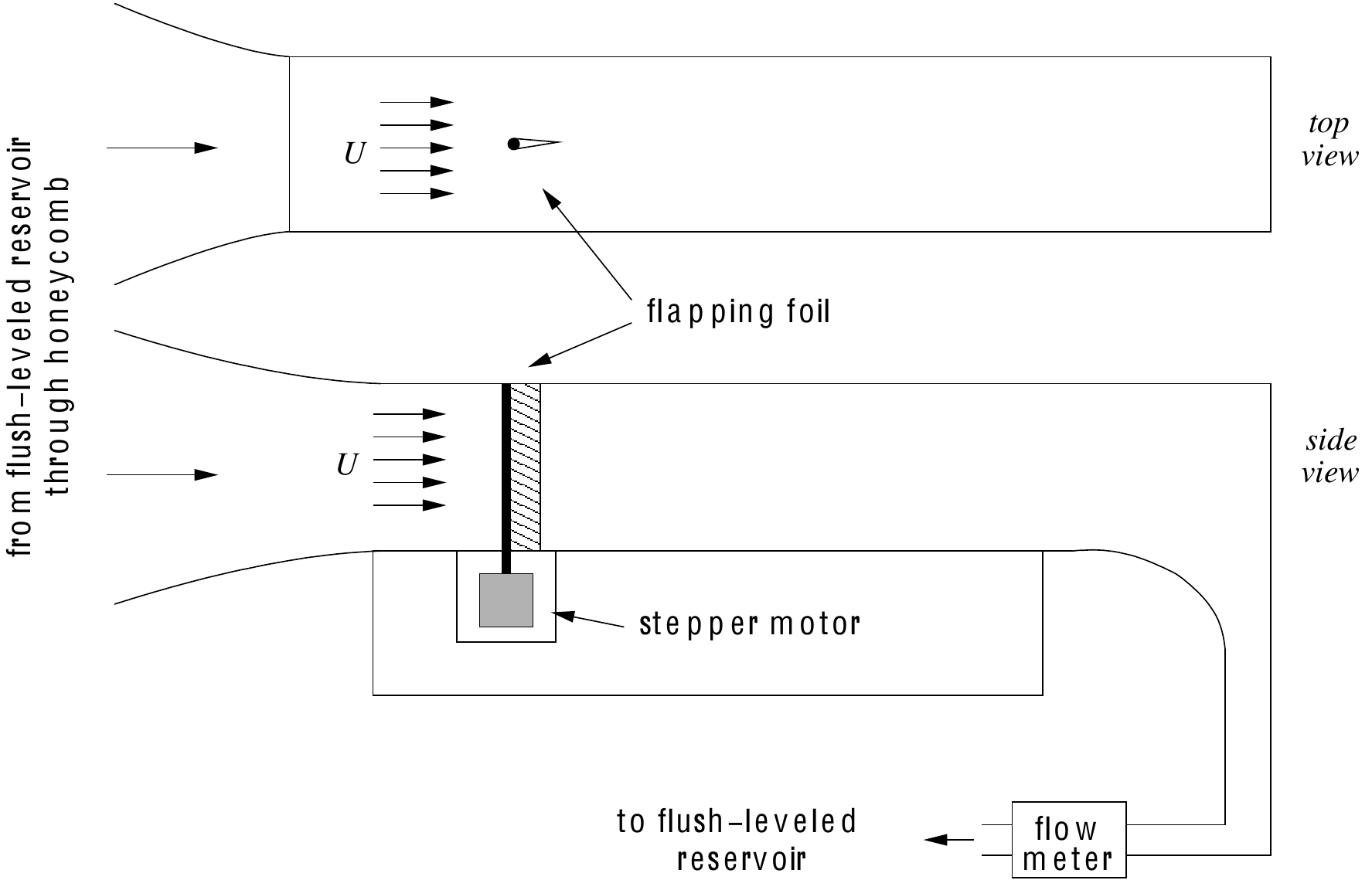}
\includegraphics[height=0.41\linewidth]{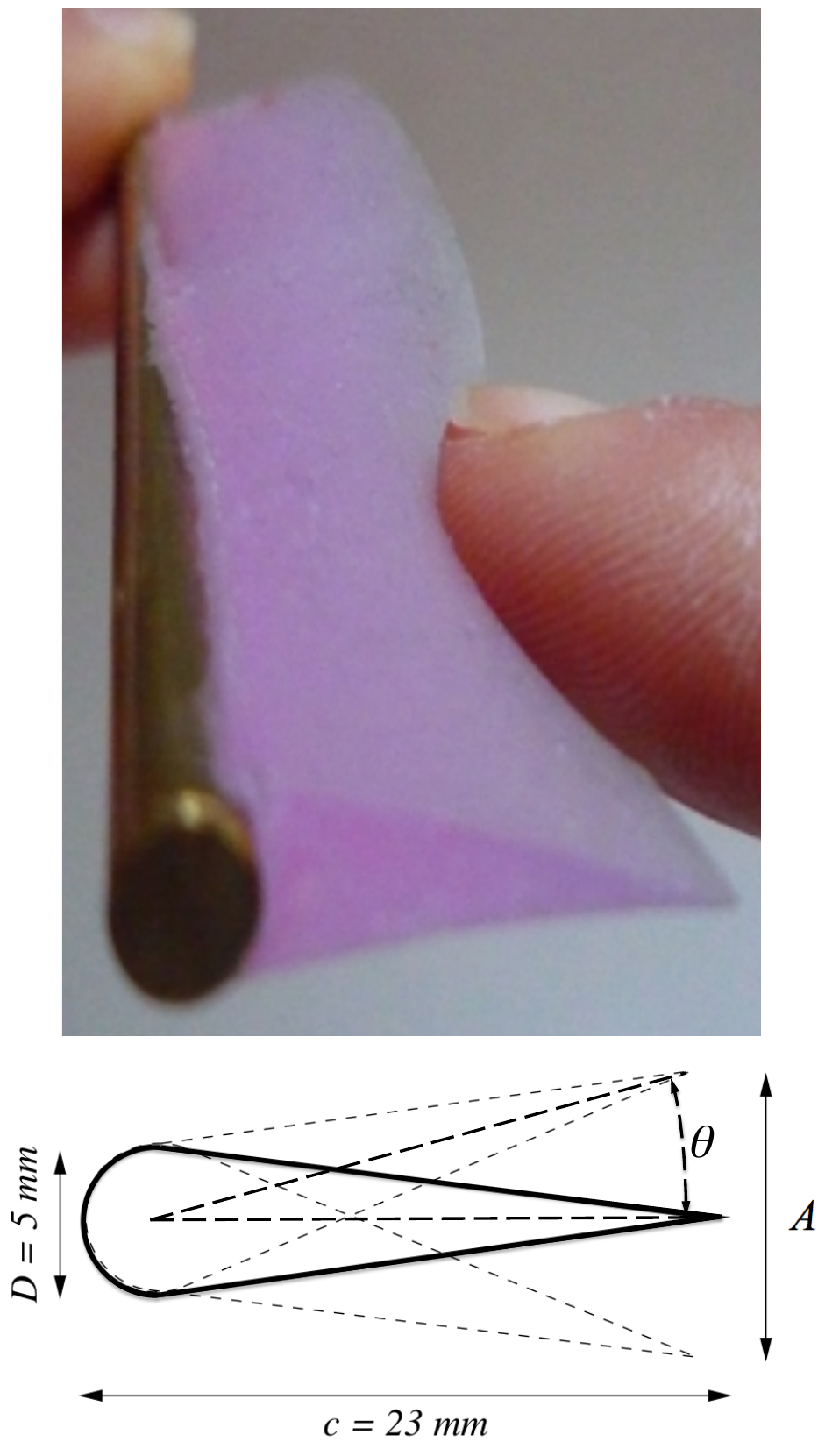}
\caption{Left: Top and side schematic views of the high-aspect-ratio
pitching foil placed in the hydrodynamic tunnel. Right: photo of the flexible foil and schematic diagram.} \label{fig_setup}
\end{figure}

\section{Experimental setup and parameters}

The experimental set-up closely resembles the one described in \cite[][]{godoy-diana2008} and
consists of a pitching foil of span-to-chord aspect ratio of 4:1
placed in a hydrodynamic tunnel (see figure \ref{fig_setup}). The motion of the foil is
driven by a stepper motor and has been set to a triangular wave of varying frequencies and amplitudes.
The control parameters on the experiment are the flow velocity
in the tunnel $(U)$ and the foil oscillation frequency $(f)$ and
pitching angle $(\theta)$. The latter determines a peak-to-peak amplitude of oscillation of the trailing edge that we define as $A$ in the case of the rigid foil. As will be discussed below, the actual displacement of the trailing edge $A_{eff}$ equals $A$ for the rigid foil but varies depending on the frequency for the flexible foil. We define the following
non-dimensional parameters: the Reynolds number $\Rey=Uc/\nu$
based on the foil chord $c$, where $\nu$ is the kinematic
viscosity, fixed to  $\Rey\approx 1035$ in the present experiments. The dimensionless effective flapping amplitude
$A_D=A_{eff}/D$, and a Strouhal number $St_D=fD/U$, both defined using the foil width $D$ as characteristic length scale. It has been shown recently
\cite[][]{godoy-diana2008} that this three-parameter description
is needed to allow for a proper characterization of the transitions
in the wake of a flapping foil. The amplitude-based Strouhal number
$St_A=St_D\times A_D$ that is usually found in the literature was in
the range $0.05<St_A<1.2$ in the present experiments.

Time-resolved measurements were performed using
two-dimensional particle image velocimetry (PIV) on a plane probing the
flow around the foil at mid-span. A continuous wave laser (DPSS 2W@532nm) and a cylindrical lens were
 used to produced a light sheet of $\approx$1mm width in the whole imaging region ($-10.4D$ to $26.6D$ and $\pm10.8D$
in the $x$ and $y$ directions, respectively). Each recording consisted of 2100 images sampled at 500 Hz using a Phantom V9.1 camera with 1600$\times$1200 pixels resolution. The flow was seeded with 20$\mu$m polyamide particles.
The PIV computation and post-processing were done using a LaVision system as well as Matlab and the PIVMat Toolbox \cite[][]{pivmat}. The PIV calculation was performed with 16$\times$16 pixel interrogation windows with 50\% overlap, giving a resolution of one vector every 0.7mm.

We investigate two cases, using foils with the same geometry and dimensions (indicated in
figure \ref{fig_setup}), but differing in their structural material.
The first one, already used in previous studies, is made of acrylic glass
and will be referred to as the \emph{rigid foil} in the following. The second
one, the \emph{flexible foil} shown in figure \ref{fig_setup}, is made of a silicone
elastomer material (vinyl polysiloxane), crosslinked with a curing agent to obtain
a Young's modulus of $E\approx30\mathrm{kPa}$.

% FIGURE 2
\begin{figure}
\centering
\includegraphics[height=0.37\linewidth]{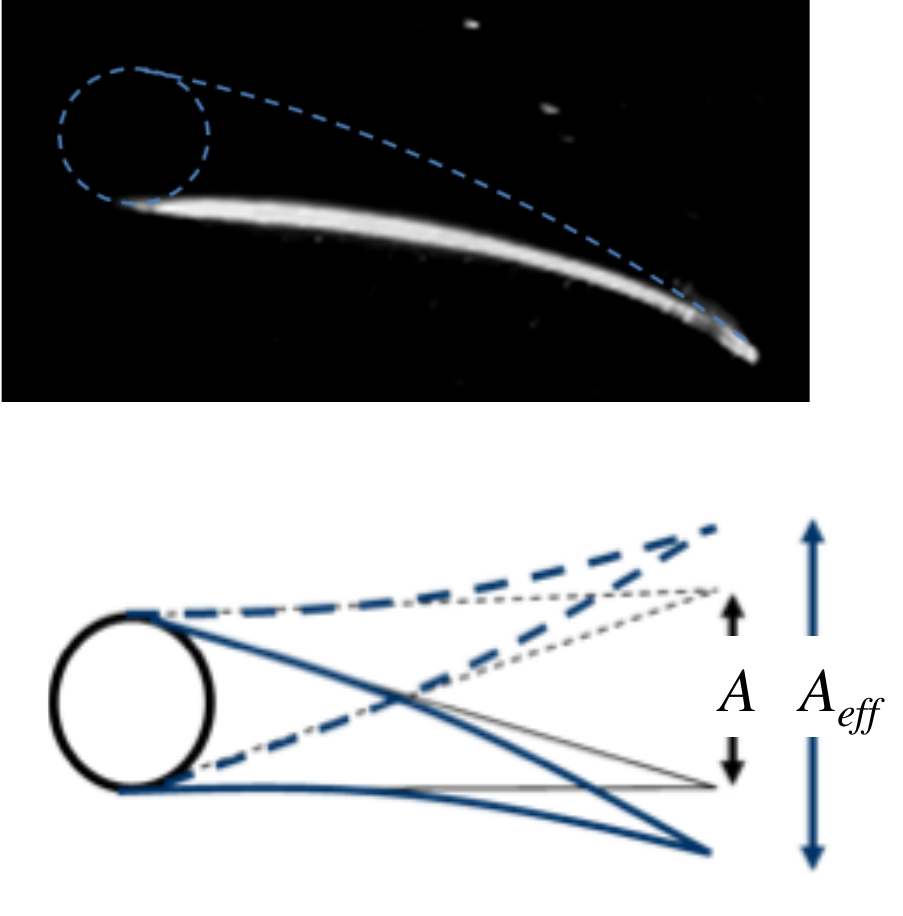}
\includegraphics[height=0.38\linewidth]{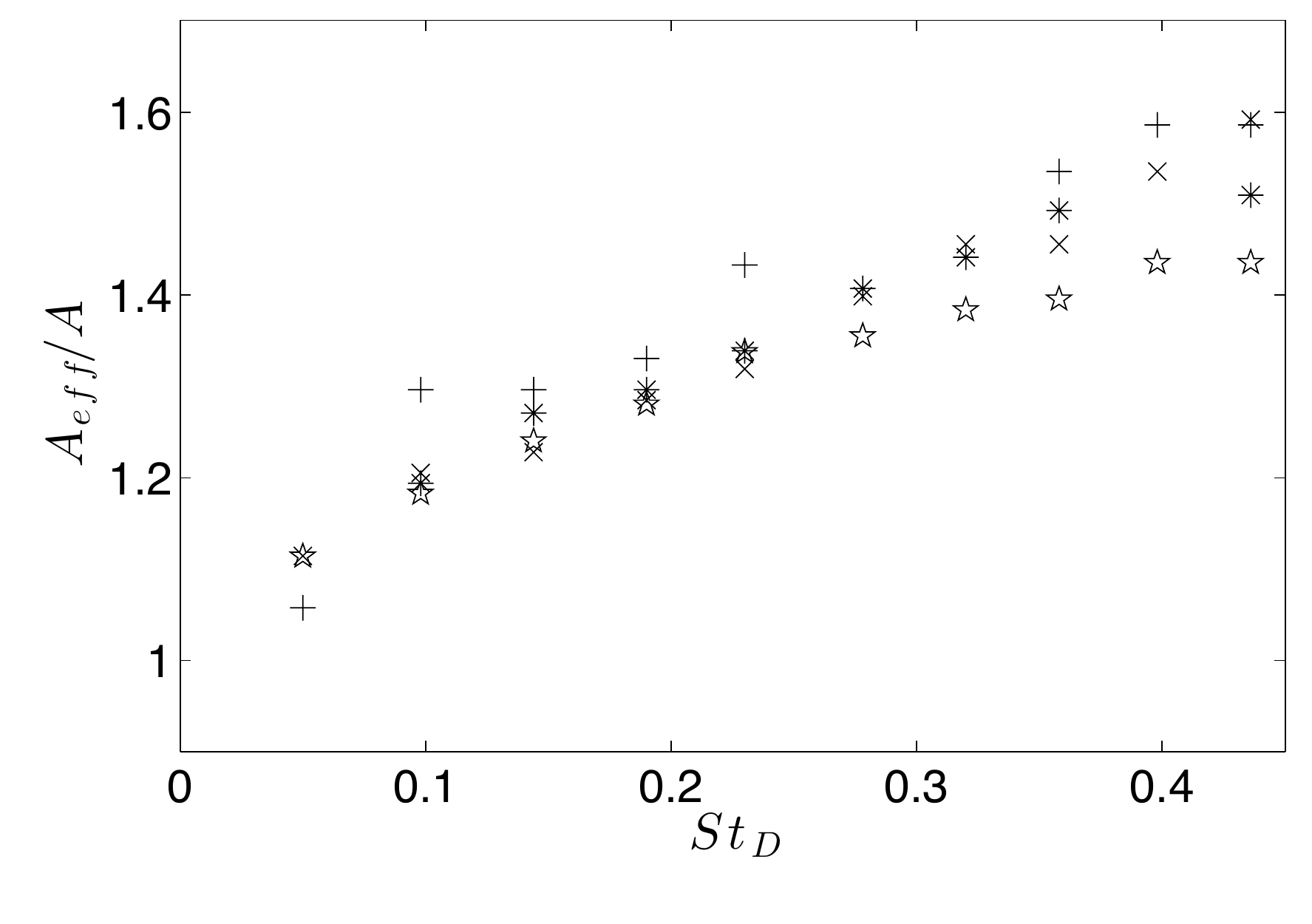}
\caption{Left: Photo of the flexible flap where the reflection of a light sheet shows the deformation of the foil profile (top) and schematic diagram defining $A_{eff}$ (bottom). Right: Effective flapping amplitude $A_{eff}$ as a function of the
Strouhal number $St_D$ for $\theta=$5$^{\circ}$ ($+$), 7.5$^{\circ}$ ($\times$), 10$^{\circ}$ ($\ast$) and 15$^{\circ}$ ($\star$) (which means $A_D=$0.7 ($+$), 1.1 ($\times$), 1.4 ($\ast$) and 2.1 ($\star$) for the rigid foil).}
\label{Aeff_st}
\end{figure}

The deformation experienced by the flexible foil while flapping can be
characterized using the actual amplitude $A_{eff}$ (peak to peak amplitude measured at the
trailing edge of the foil), which is larger than the amplitude $A$ that
would be observed with a rigid foil for a given angular
oscillation.  The ratio $A_{eff}/A$ thus increases with the flapping
frequency for a given value of the imposed rigid amplitude $A$ (see
figure \ref{Aeff_st}, where a schematic diagram of this effect is also shown). The latter observation holds for the regime considered in the present study, however one could likely find other regimes where $A_{eff}/A<1$. This is related to the problem of an optimal flexibility \cite[see e.g.][]{michelin2009} as a function of the forcing frequency which, although beyond the scope of the present work, remains to be explored experimentally.

\section{Transitions in the wake}

\begin{figure}
\centering
\includegraphics[width=0.48\linewidth]{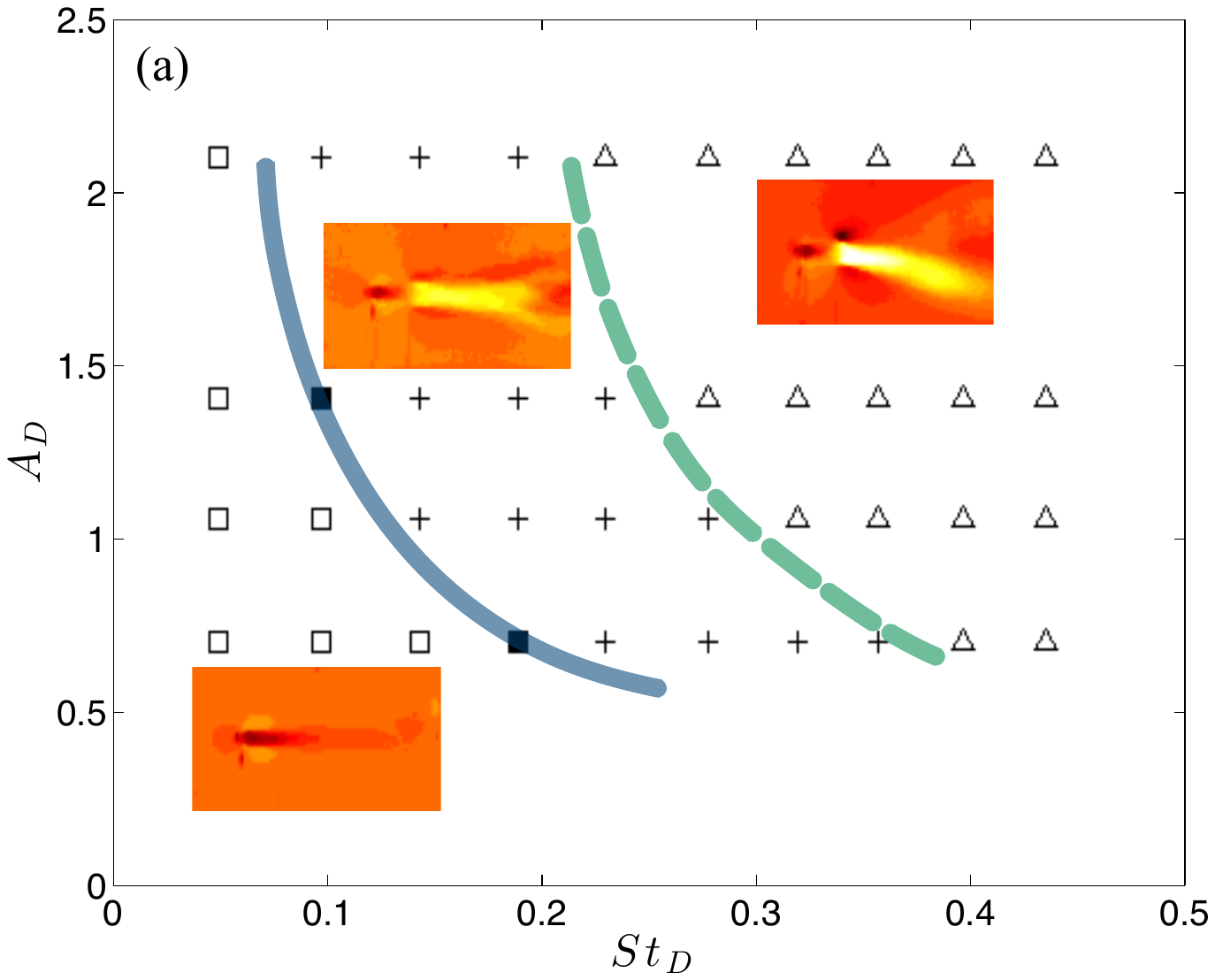}
\includegraphics[width=0.48\linewidth]{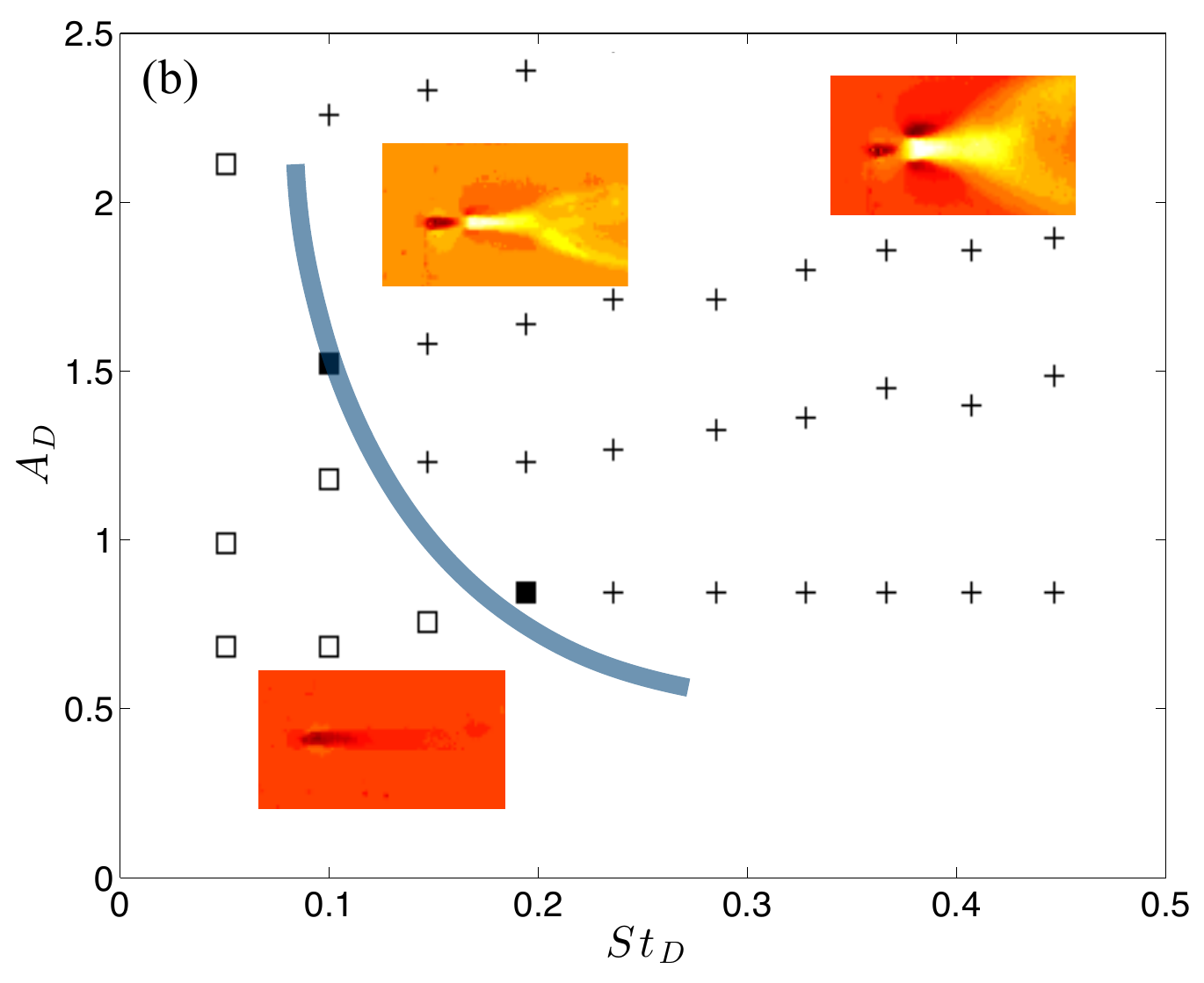}
\includegraphics[width=0.96\linewidth]{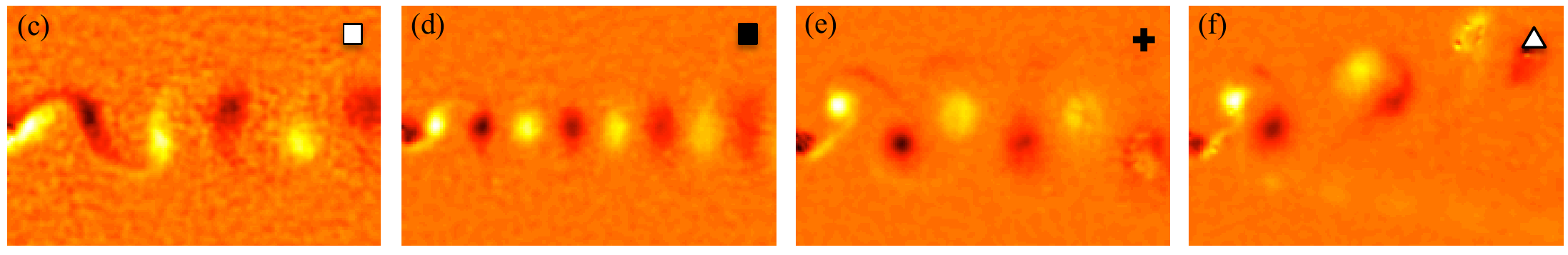}
\caption{Amplitude vs. Frequency phase diagram for the transitions in the wake for (a) the rigid foil \cite[see also][]{godoy-diana2009} and (b) the flexible foil. Inserted images correspond to a typical mean horizontal velocity field for each regime. Solid and
dashed lines correspond to transitions between regimes (see text). For the flexible case, the transition that marks the symmetry breaking
of the reverse B\'enard-von K\'arm\'an wake is not observed. (c-f) Snapshots of typical vorticity fields for each regime.}
\label{PhaseSpace}
\end{figure}

A description using the frequency and amplitude of the oscillatory motion
as independent parameters has been shown recently to be the optimum framework to
fully characterize the quasi-two-dimensional (Q2D) regimes observed in the wake of a
pitching foil \cite[][]{godoy-diana2008,schnipper2009}. The phase diagram indicating the regime
transitions for the rigid foil in a $(St_D,A_D)$ map is recalled in Fig.~\ref{PhaseSpace}(a), where a typical snapshot of the mean horizontal velocity
field is shown for each region of the phase space (snapshots of a typical vorticity field for each case are shown below in Fig.~\ref{PhaseSpace}c-f). The solid line marks the transition between a B\'enard-von K\'arm\'an type (BvK) wake (where the mean flow presents a velocity deficit behind the foil) and the reverse BvK wake characteristic of propulsive regimes (where the mean flow is a jet). The broken line locates the symmetry breaking of the reverse BvK street that occurs when two successive vortices are shed too close to each other,
the closeness being determined by a ratio of the flapping and advective timescales \cite[see][]{godoy-diana2009}.

% FIGURE 3

\begin{figure}
\centering
\includegraphics[width=\linewidth]{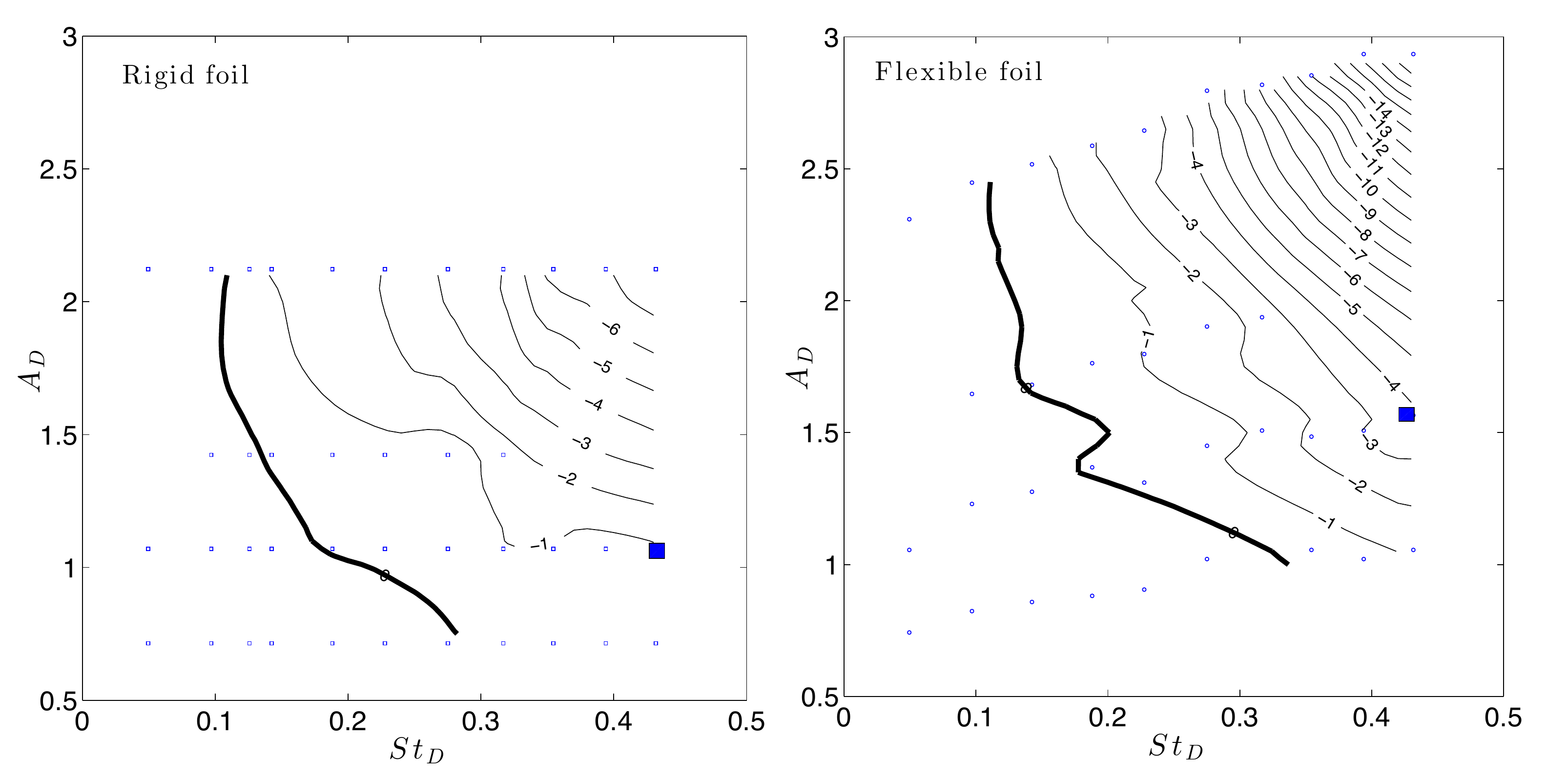}
\includegraphics[width=\linewidth]{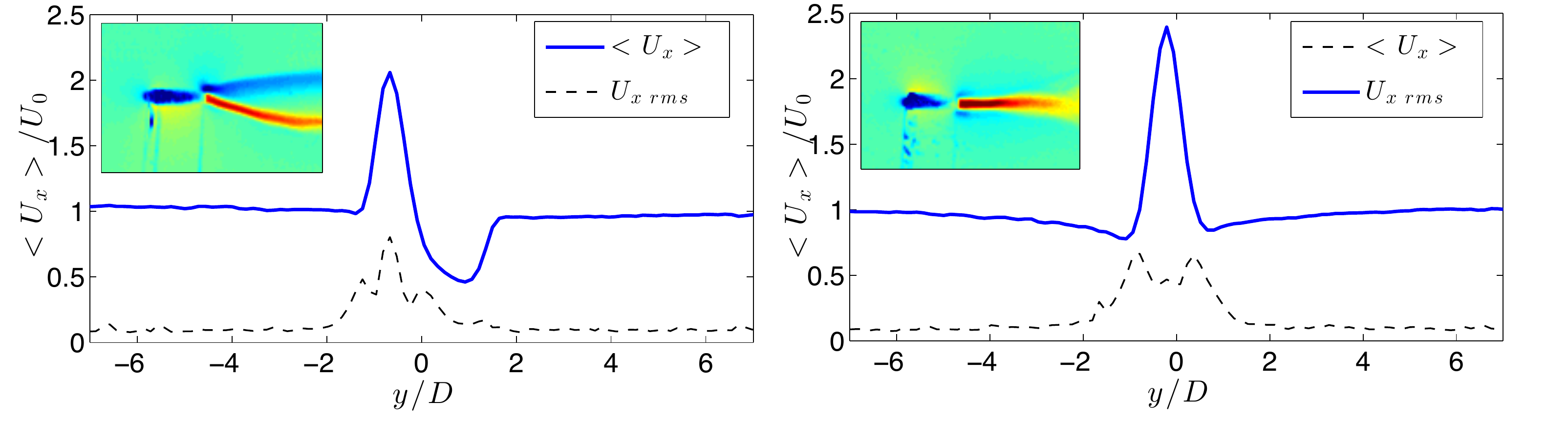}
\caption{Top row: Isocontours of a mean drag coefficient $C_D/C_{D0}$ surface
estimated using a momentum-balance approach. The black contour corresponds to $C_D=0$ where the estimated drag-thrust transition occurs \cite[see][]{godoy-diana2008}. Bottom row: Cross-stream ($y$) profiles of the mean horizontal velocity on the downstream boundary of the control volume used for the thrust force estimation (solid lines) and corresponding rms (dashed lines) for a typical case (marked with a square symbol in the top plots) where the symmetry breaking is inhibited by the effect of foil flexibility. Insets: Snapshots of the mean $<U_x>$ field showing the jet flow in the wake of the flapping foil.}
\label{force}
\end{figure}

The effect of adding flexibility to the flapping foil modifies the structure of the flow in the wake and thus the transition map (see Fig.~\ref{PhaseSpace}b). While there is no observable effect in the transition between the BvK and reverse BvK regimes
(i.e. the solid line does not move appreciably between Figs. \ref{PhaseSpace}a and \ref{PhaseSpace}b),
the symmetry breaking of the BvK street is inhibited for the flexible foil in the range
of parameters explored. The propulsive jet that is deflected for the rigid foil (in the
top-right region of the $(St_D,A_D)$ parameter space in Fig. \ref{PhaseSpace}a) stays symmetric in the
case of the flexible foil. This can be observed in more detail in the comparison of the mean horizontal velocity profiles for a specific point of the parameter space shown in Fig. \ref{force}.

A standard momentum balance using the mean velocity field  permits to estimate
the average thrust force \cite[see][]{godoy-diana2008}. Fig. \ref{force}
shows the mean drag coefficient $C_D/C_{D0}$ surfaces in the $(St_D,A_D)$ map
for the two foils\footnote{As mentioned in \cite{godoy-diana2008}, the drag-thrust transition (the solid contour
where $C_D=0$) happens \emph{after} the transition from BvK to reverse BvK
regime.  This observation has been confirmed by \cite{bohl2009} who used an alternative control volume analysis that takes into account the streamwise velocity fluctuations and the pressure term.}. We recall that for a given angular amplitude $\theta$ of the pitching motion, at equal flapping frequency the effective amplitude has been observed in the present experiments to be larger for the flexible foil than for the rigid one. Now if we compare two points in the parameter space at equal effective amplitudes (i.e. equal $A_D$), the calculated forces for the flexible and the rigid foils are indeed similar (see contours in Fig. \ref{force}). Seeking to characterize the effect of flexibility however, one should consider the pitching motion as the input and the passive deformation of the foil as an effect, which would mean comparing points of equal angular amplitude $\theta$ (e.g. the square marker in Fig. \ref{force}) where the propulsive force is significantly larger for the flexible foil. Comparing forces between the rigid and the flexible foils at equal values of angular pitching amplitude, we can see that the thrust force for the flexible foil can be up to three times larger than the values obtained with the rigid foil, a feature that seems to be mainly determined by the increase in effective flapping amplitude resulting from the foil deformation.

% FIGURE 4

The present experiments have thus evidenced that foil flexibility has an effect on both the thrust force produced by the foil and the symmetry properties of the wake. In what follows we attempt to gain a better understanding of the mechanisms responsible for these
observations focusing on the formation of vortices in the case of the reverse BvK wake,
where the vorticity generated at the leading edge and along the boundary layer on each side
of the flap merges with the tip vortex at the trailing edge and is shed
as a single structure each half period of oscillation. The interest in the early stages of the vortex formation process is two-fold: on the one hand, because vortices are formed in the vicinity of the oscillating foil, it is during these first stages that their influence or footprint on the production of hydrodynamic forces will be greatest. Second, it has been shown that the symmetry properties of the wake are driven by the dynamics of the very near wake, where the formation of dipolar structures is the mechanism that triggers the symmetry breaking of the reverse BvK vortex street.

% FIGURE 5
\begin{figure}
\centering
\includegraphics[width=0.9\linewidth]{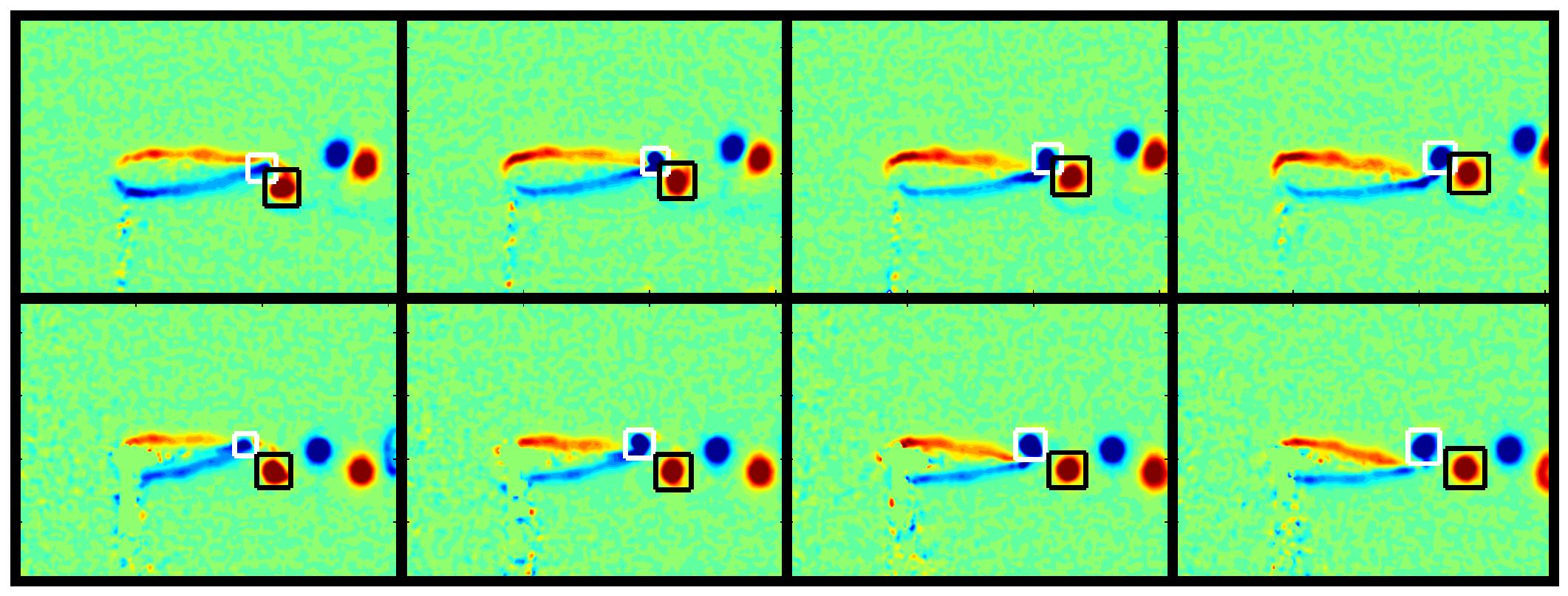}
\caption{Four successive snapshots of the vorticity field (from left to right) showing the temporal tracking of vortices in the wake of a rigid foil
(top row) and flexible foil (bottom row). The tracked structures will be labeled vortex I (tracked with the white square) and vortex II (black square), see text. The case shown corresponds to $St_D$=0.3 and $\theta=7.5^{\circ}$, the latter corresponding to $A_D$=1.1 for the rigid foil and $A_D$=1.5 for the flexible foil.} \label{SuiviTbs}
\end{figure}

% FIGURE 6
\begin{figure}
\centering
\includegraphics[width=1\linewidth]{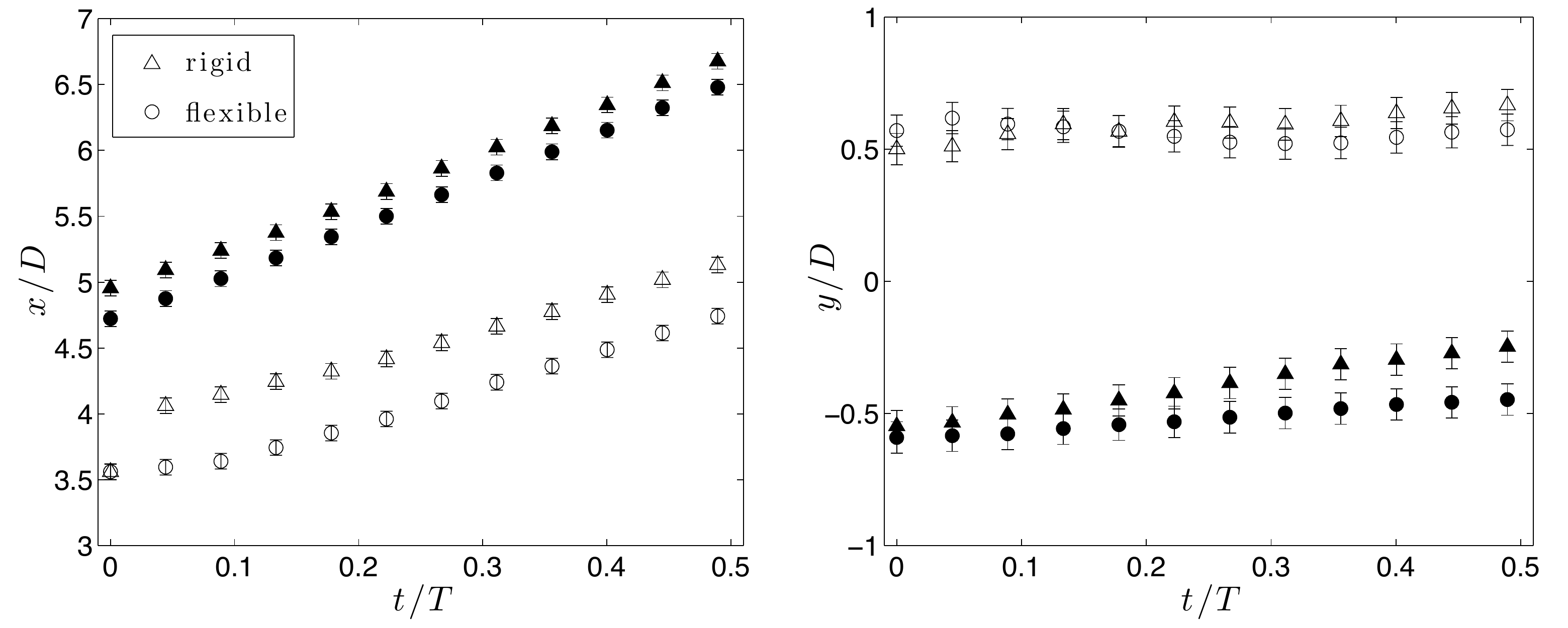}
\caption{Temporal evolution of the streamwise ($x$) and transversal ($y$)
positions of a vortex pair obtained from the tracking shown in Fig.~\ref{SuiviTbs}. The symbols distinguish the rigid
(triangles) and the flexible (circles) cases. The vortex in formation still attached to the foil (vortex I) is shown in empty symbols while the
vortex already formed, further downstream in the wake (vortex II) is shown in full symbols. Errorbars show the uncertainty in the position of the extrema of vorticity that are used to track each vortex resulting from the resolution of the vorticity field.}
\label{PositionTbs}
\end{figure}

\section{Vortex dynamics}

For each experimental configuration, we start tracking the vortex closest to
the trailing edge at the moment of reversal in the direction of the
flapping motion, which corresponds to the birth of the new vortex.
The vortex is characterized in terms of its circulation and streamwise
and transverse position in the wake as in \cite{godoy-diana2009}. In figure \ref{SuiviTbs}, a time series of vorticity fields is shown for each type of foil, rigid and flexible, for the same control parameters: frequency ($St_D$=0.3) and pitching angle ($\theta=7.5^{\circ}$). The two vortices closest to the flap (counter-rotating, enclosed by squares in Fig.~\ref{SuiviTbs}) are tracked, giving the time series of position shown in Fig.~\ref{PositionTbs}. We will refer to the vortex closest to the flap (negative vorticity) as vortex I and the one shed previously (positive vorticity) as vortex II. 
The first point concerns the time evolution of the streamwise position of the vortices: we observed systematically that in the case of the flexible foil, vortices are shed closer to the origin (the rotation axis of the foil) in the $x$ direction than in the case of the rigid foil (this can be seen in the the track of vortex I in Fig.~\ref{PositionTbs}). The difference between the two foils is less marked in the track of vortex II.  The tracks of the $y$ position (Fig.~\ref{PositionTbs}) show the wake deflection observed for the rigid foil in this case, while in the case of the flexible foil the vortices are shed symmetrically.

\begin{figure}
\centering
\includegraphics[width=0.6\linewidth]{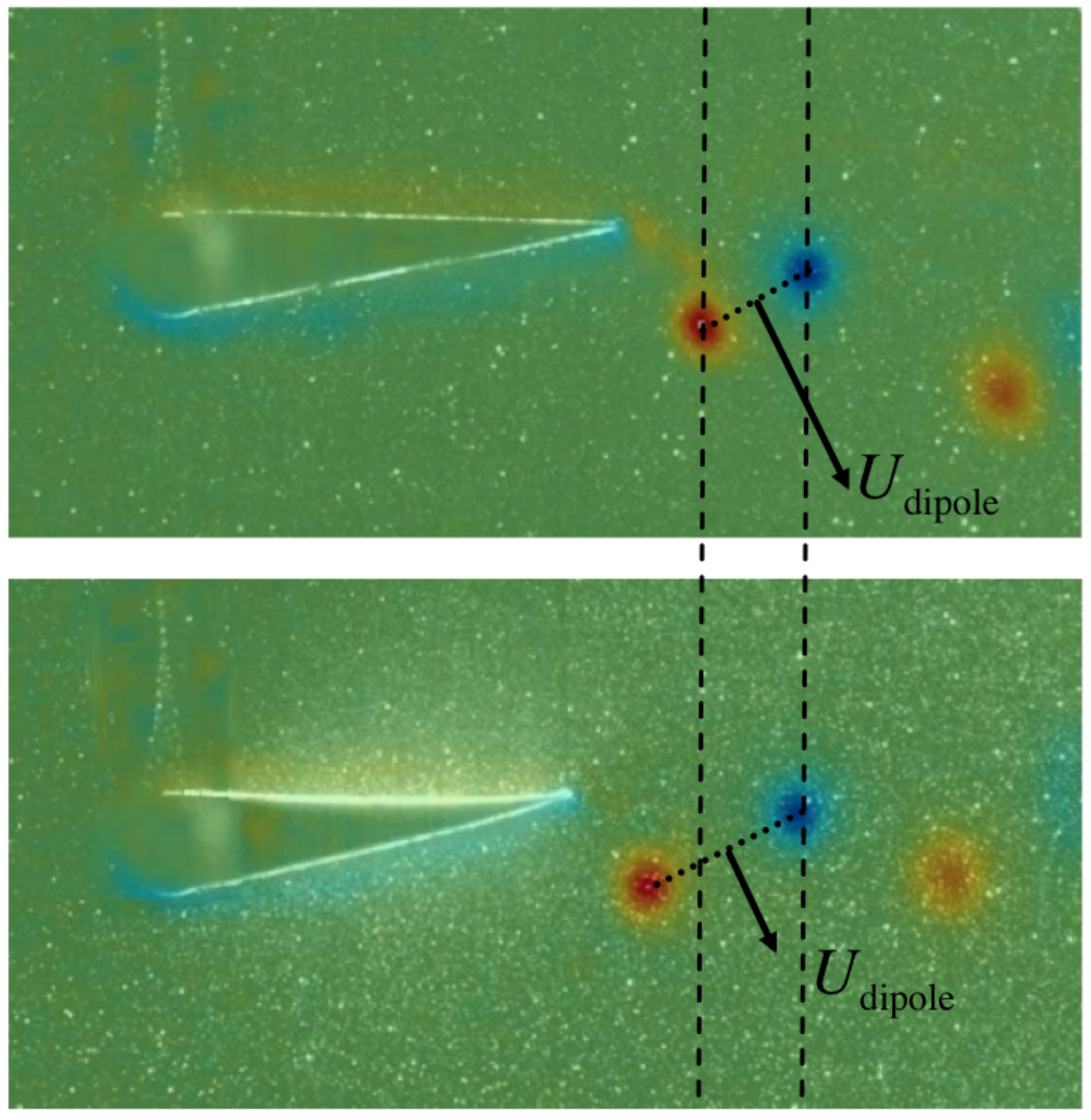}
\caption{Vorticity field in the near wake for (top) the rigid foil and (bottom) the flexible foil. The dashed lines mark the horizontal distance between the centers of two consecutive vortices in the rigid case, pointing out the larger distance between the two vortices in the flexible case.  The self-advection velocity of the dipolar structure $U_{\mathrm{dipole}}$ is shown schematically in both cases, larger in the rigid case where the dipole is more compact.}
\label{SchemaTbs}
\end{figure}

The difference between the $x$ position of vortex I in the rigid and flexible cases
can be understood considering the deformation experienced by the flexible
foil during the vortex formation process: when the flap changes the direction of motion, because of the foil deformation the tip position moves closer to the origin and so does the shed vortex. It follows from this observation that the two consecutive vortices of opposite
sign shed at each flapping period, are closer in the rigid case
than in the flexible case, as illustrated in Fig.~\ref{SchemaTbs}. Recalling that the symmetry-breaking in the wake directly depends on the
relative self-advection velocity of the first dipolar structure in
the near wake of the foil \cite[][]{godoy-diana2009}, the fact that flexibility keeps the newly shed vortex (vortex I) from being shed close enough to vortex II to form a vortex pair can explain the observed inhibition of the deflected wake.

% FIGURE 7

\begin{figure}
\centering
\includegraphics[width=0.7\linewidth]{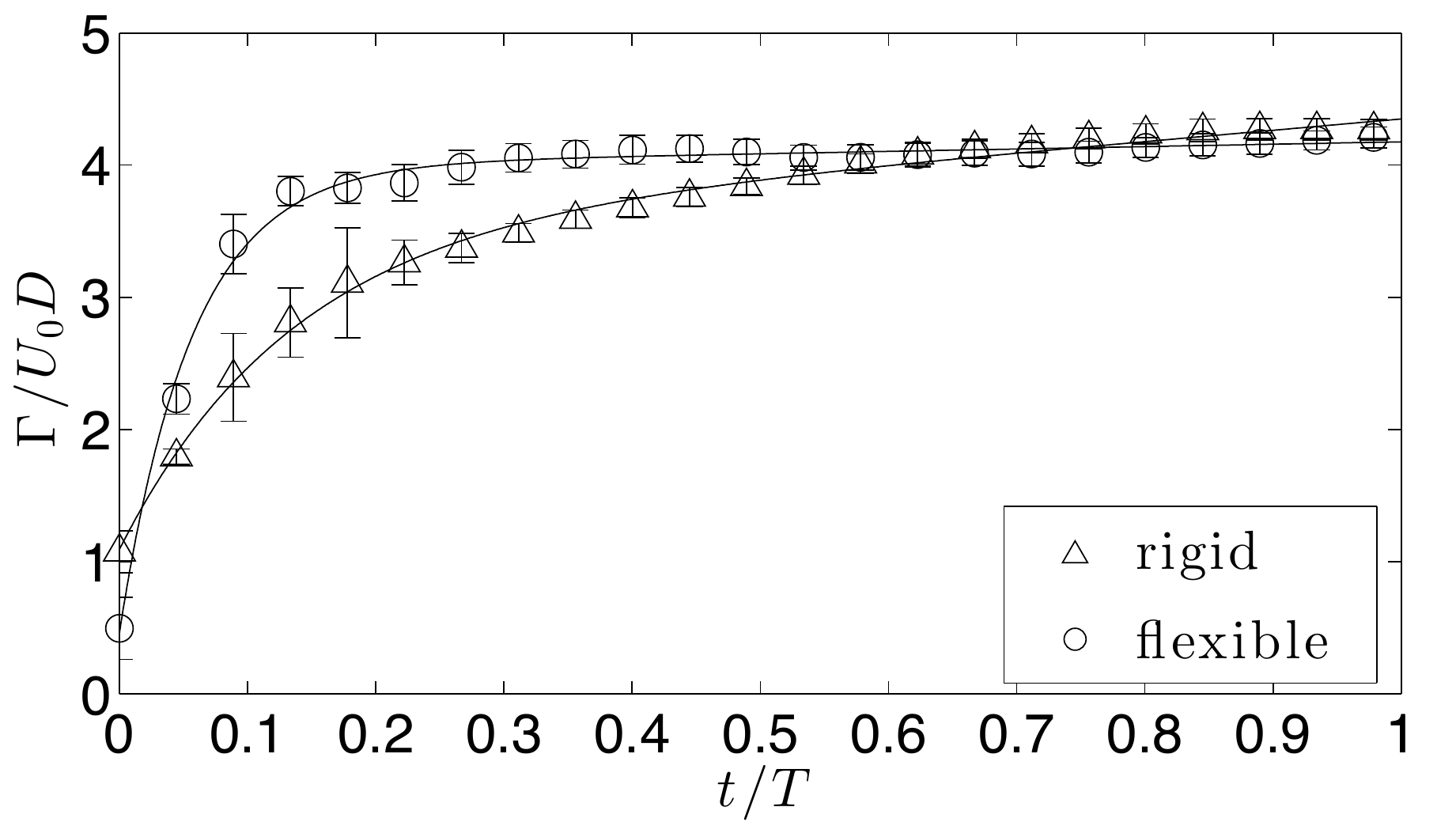}
\caption{Temporal evolution of the vortex circulation for the same cases of Fig.~\ref{SuiviTbs}. Errorbars show the difference between two methods of calculating the circulation using either the velocity or the vorticity fields.} \label{CircTbs}
\end{figure}

We study now the temporal evolution of the circulation during vortex formation. 
We define the time origin as the moment when the foil changes direction, which triggers the shedding of
a vortex into the wake and marks the beginning of the growth process of a new vortex at the trailing edge of the foil.
The circulation grows in time and rapidly tends toward a maximum value that remains relatively constant even after the next change in direction (which restarts the formation process for the next opposite-signed vortex).
This evolution can be seen in Fig.~\ref{CircTbs}. If we compare this circulation evolution for flexible
and rigid foils, it appears clearly that the vortex in the wake of
the flexible foil forms faster than in the case of the rigid foil.

The link between vortex formation time and the force experienced by the foil can be qualitatively understood using the simple approach proposed by \cite[][]{protas2003} for a 2D wake vortex street, starting from the impulse formula that associates the force with the time derivative of the vorticity impulse \cite[see e.g.][]{saffman} $\vec{F} = -\frac{\mathrm{d}}{\mathrm{d}t}\int_{\Omega} (\vec{r} \times\vec{\omega}) \mathrm{d}\Omega$, where the fluid density is set to unity, the flow domain $\Omega$ extends to infinity, $\vec r = [x,y]$ is the position vector, $\vec{\omega} = (\partial_x v -\partial_y u)\vec k$ the vorticity, $u$ and $v$ are the two velocity components, and $\vec k$ the unit vector perpendicular to the plane of motion. Considering a vortex street modeled by point vortices, the projection of the impulse formula along the $x$ axis, gives the following approximate expression for the drag force : $F_D \cong -\frac{\mathrm{d}}{\mathrm{d}t}\sum_{i} \Gamma _i \Delta y_i$,  where $i$ denotes the $i$-th vortex in the wake and $\Delta y_i$ its distance to the wake centerline. The hydrodynamic force is thus on the one hand intimately linked to the spatial arrangement  of vortices in the wake, as well as to their sign and intensity. The other important point is that this unsteady contribution to the force coming from the vortices depends on the temporal evolution of the circulation. Considering the comparison between the rigid and flexible foils of the present experiments, the faster formation of each vortex in the case of the flexible foil is therefore consistent with the observed increase in the propulsive force.

% FIGURE 8

\begin{figure}
\centering
\includegraphics[height=0.45\linewidth]{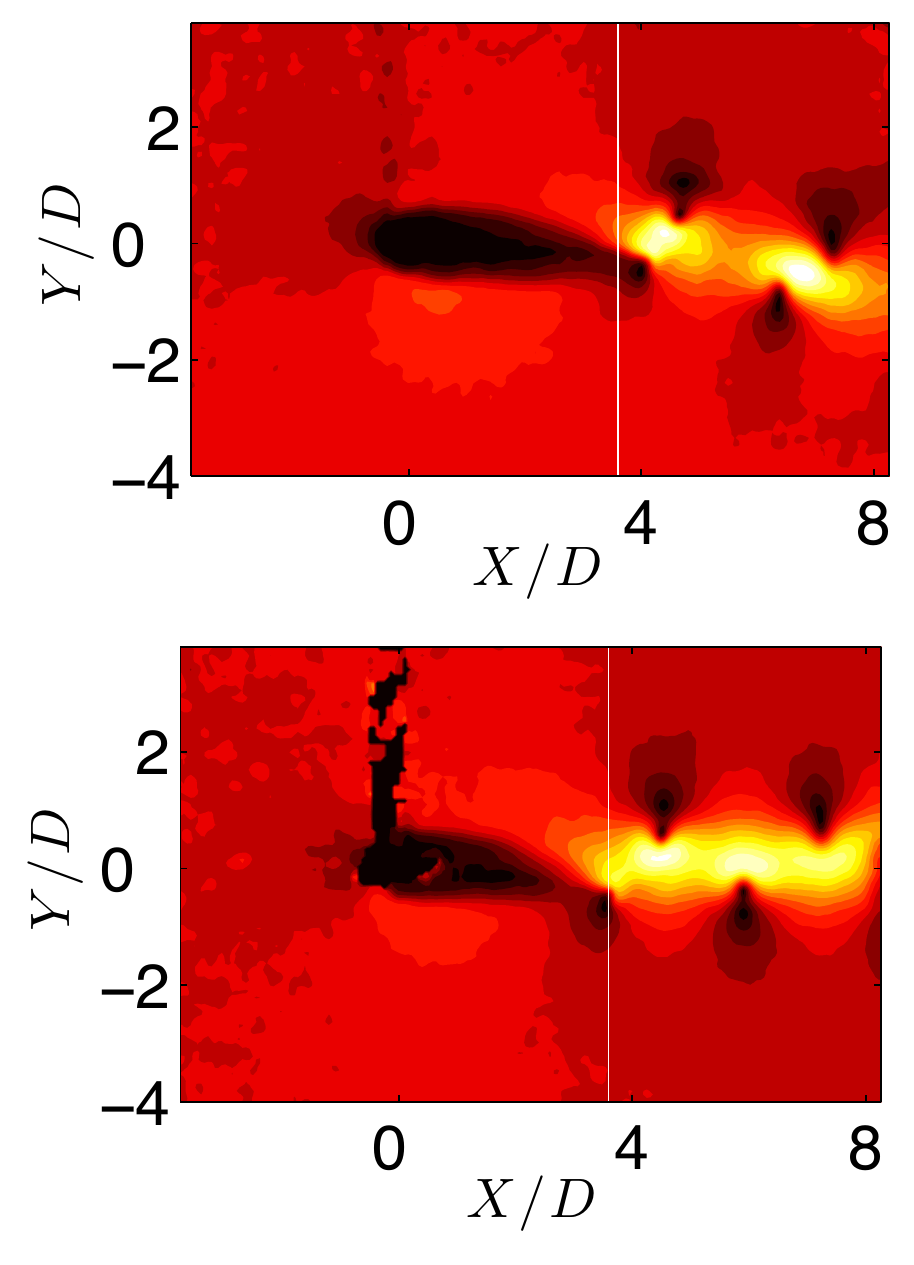}
\includegraphics[height=0.48\linewidth]{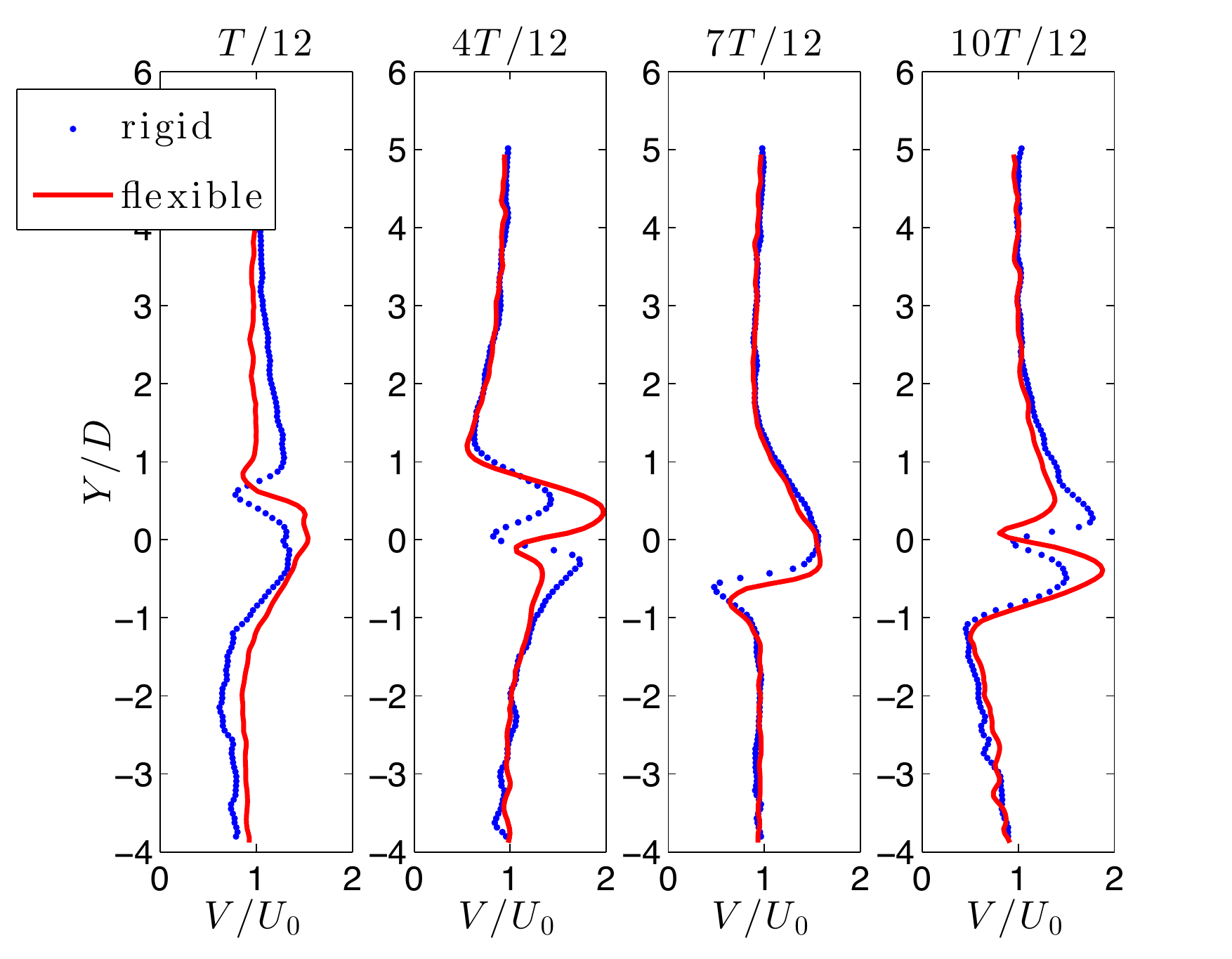}
  \caption{Comparison of the near wake velocity field for the flexible and rigid foils.
  Left: Modulus of the velocity fields (phase averaged, at T/12) for rigid foil (top) and flexible foil (bottom).  Right: Profiles taken along the white line marked over the contour plots on the left hand side (phase averaged, at 4 different times).}
    \label{fig_comparaison}
\end{figure}

In order to shed some light on the physical mechanism that determines the faster formation of vortices
in the wake of the flexible foil we examine the velocity field in the near wake. We compare in
 Fig.~ \ref{fig_comparaison}, the velocity profile $|V(t)|$ at
the trailing edge of the foil for both cases (along the white
line represented on the absolute velocity fields in Fig.~ \ref{fig_comparaison}, left),
at four different instants distributed over one foil oscillation period.
The profiles for the rigid and flexible foils are similar, but a distinctive feature appears when the foil changes direction of oscillation ($\frac{4}{12} T$ and $\frac{10}{12} T$): the velocity jet near the tip of the foil is clearly more intense in the case of the flexible foil. The faster growth of the vortex circulation observed in Fig.~\ref{CircTbs} can therefore be explained acknowledging that the flexible foil curvature enhances the flow around the trailing edge of the foil, locally accelerating the fluid that will feed the growing vortex.

% FIGURE 9

We may also note that this is compatible with another point observed in the
previous section: in the flexible case, once the newly formed vortex is released into the wake it accelerates and catches up its
delay with respect to the case of the rigid foil, (see the $x$ position of vortex II for both cases in Fig.~\ref{PositionTbs}). This transient faster initial advection velocity of the vortices can be explained by the deformation dynamics of the flexible structure:
in the manner of a spring, after reaching its maximum deformation the foil structure comes back to its initial
shape, contributing to accelerating the fluid in the near wake. 

\

\section{Concluding remarks}

We have shown that adding flexibility plays a strong dynamical role on the
wake produced by a flapping foil.
On the one hand, the effective
amplitude obtained passively due to the deformation of the flexible foil while
flapping (Fig.~\ref{Aeff_st}) leads to an increase in the propulsive force
with respect to the case of the rigid foil (Fig.~\ref{force}). 
On the other hand, the interaction of the shed vortices
with the flexible structure inhibits the trigger of the symmetry
breaking of the reverse B\'enard-von K\'arm\'an wake (Fig.~\ref{PhaseSpace}b), neutralizing thus the deflection of the propulsive jet that has been widely reported in the literature. The latter result, which is a novel observation, brings yet more evidence that wing compliance needs to be considered as a key parameter in the design of future flapping-propelled vehicles: since not only it is determinant for thrust and efficiency, but also as we show here because of its role in dictating the vortex dynamics that governs the stability properties of the wake.

\begin{acknowledgments}

We thank gratefully D. Pradal and G. Clermont for their valuable help manufacturing the flexible foils and J. Bico for useful discussions. We acknowledge support  from the French National Research Agency through project ANR-08-BLAN-0099.

\end{acknowledgments}

\end{document}